# Carbon in the trigonal prismatic environment of rheniums in $[Re_{12}CS_{17}(CN)_6]^{n-}$ complexes

Svyatoslav P. Gabuda,[†] Svetlana G. Kozlova,*[†‡] Vladimir A. Slepkov,[†]
Yuri V. Mironov[†], Vladimir E. Fedorov[†]

*†Nikolaev Institute of Inorganic Chemistry, SB RAS, 630090 Novosibirsk, Russia*
*‡Boreskov Institute of Catalysis, SB RAS, 630090 Novosibirsk, Russia*

E-mail sgk@che.nsk.su

The metal cluster compounds are generally regarded as promising subjects for molecular electronics and for construction of nano-size materials. A recently discovered molecular species $[W_6CCl_{18}]^{n-}$ (n=0–3) provides some prospects of utilizing this totally special type of cluster unit [1,2]. The structure of the clusters is a carbon-centered $W_6$ trigonal prism augmented by twelve edge-bridging chloride anions and six radial chloride ligands. A lately synthesized sulphide-cyanide rhenium complex $[Re_{12}CS_{17}(CN)_6]^{n-}$ (n = 6, 8) includes carbon $\mu_6$-C in trigonal prismatic environment of 6 rheniums. The unusual behaviour of the complexes is a sharp change of interatomic distances in the vicinity of $\mu_6$-C when the oxidation state of the complexes changes (Fig. 1) [3]. We analyze the electronic density of the complexes with different oxidation states by the method of electron localization function (ELF) to study the interactions of the central carbon atom $\mu_6$-C with its surrounding.

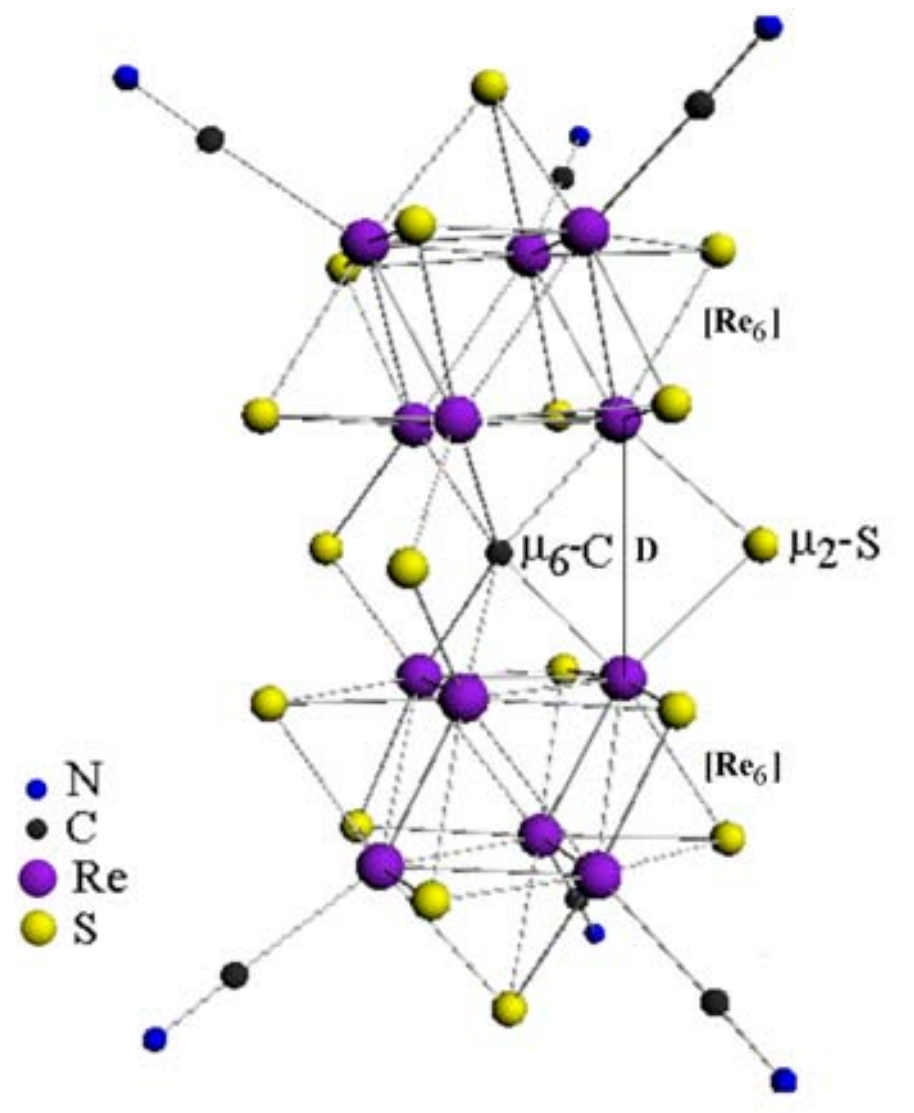

Figure 1. Structure of $[Re_{12}CS_{17}(CN)_6]^{n-}$ (n = 6, 8) clusters[3]. The distance (D) between rheniums is 2.90 Å for n = 6 and 3.17 Å for n = 8.

We studied the electronic structure of $D_{3h}$ symmetry model systems $[Re_{12}CS_{17}(CN)_6]^{n-}$ (n = 6, 8) by spin-restricted DFT method using the ADF2006 program package [4]. The model Hamiltonians included the local density functional LDA (VWN [5]) and the gradient exchange functional GGA (Becke [6]& Perdew [7]). Basis functions for all atoms consisted of the TZP/ADF2006 Slater relativistic valence triple zeta plus 1 polarization function. Geometry optimization was performed for $D_{3h}$ point group, relativistic effects were taken into account with the zero order relativistic approximation method ZORA [8]. Interactions between $\mu_6$-C and its surrounding were studied with the topological method of electron localization function ELF [4,9]. The ELF function is an orbital independent description of the electron localization given by

$$\eta=1/(1+(D_\sigma/ D^0_\sigma)^2)$$

where $D_\sigma$ and $D^0_\sigma$ represent the curvature of the Fermi hole for the studied system and the homogeneous electron gas with the same density, respectively. ELF values are obviously confined within the interval between 0 and 1. The $\eta$ values from 1 to ~0.75 (blue color) correspond to high electron localization, lone pairs, and strong covalent bonds (bosonic behavior of the electronic density). The regions with $\eta \approx 0.5$ (green color) and $\eta << 0.5$ (from yellow to red) correspond to electron-gas-like pair probability responsible for ionic, correlation and Van der Waals interactions (unshared interactions). The chemical meaning of the function has been the subject of several interpretations [9-16].

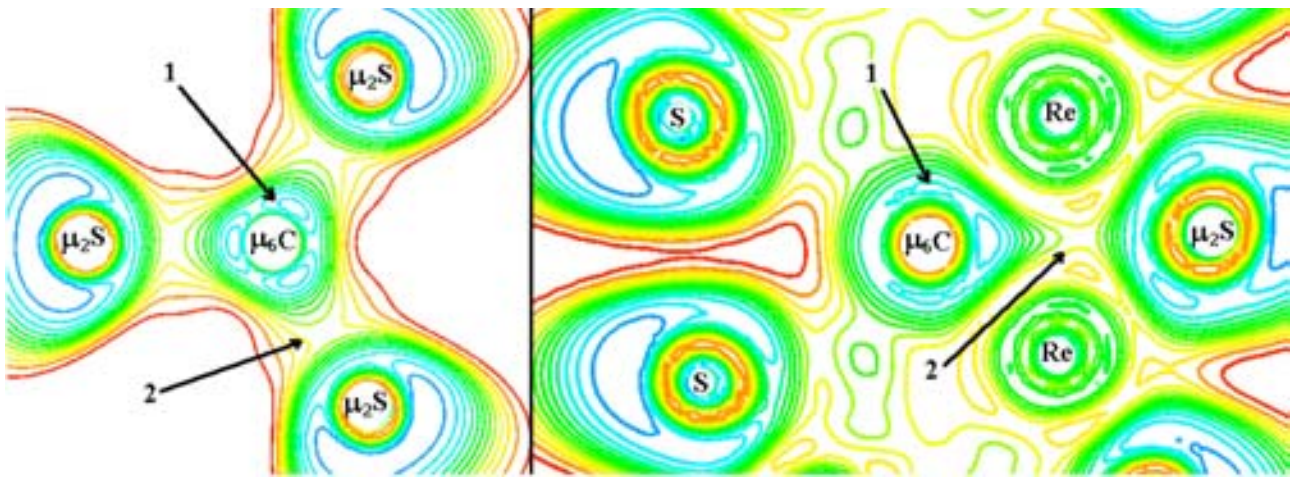

Figure 2. Distribution maps of electron density (ELF) for $[Re_6CS_{17}(CN)_6]^{6-}$ complexes: (left) the plane crossing three atoms $\mu_2$-S and the atom $\mu_6$-C; (right) the plane crossing the atoms Re- $\mu_2$S- Re- $\mu_6$C, the Re atoms belong to different fragments [$Re_6$]. Arrows 1 point to the attractors of $sp^2$-hibridization, arrows 2 point to the binding region between $\mu_6$-C and $\mu_2$-S.

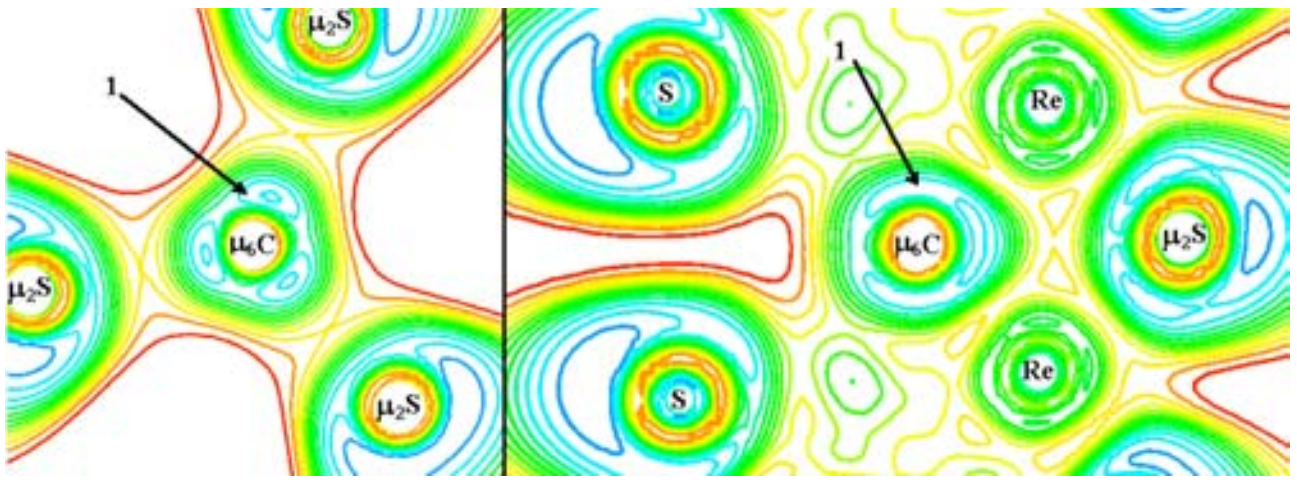

Figure 3. Distribution maps of electron density (ELF) for $[Re_6CS_{17}(CN)_6]^{8-}$ complexes: (left) the plane crossing three atoms $\mu_2$-S and the atom $\mu_6$-C; (right) the plane crossing the atoms Re- $\mu_2$S- Re- $\mu_6$C, the Re atoms belong to different fragments [$Re_6$]. Arrows 1 point to the attractors of $sp^2$-hibridization.

Fig. 2, 3, 4 show the results of the topological analysis of the electron density in $[Re_{12}CS_{17}(CN)_6]^{n-}$ (n = 6, 8) in the neighborhood of $\mu_6$-C. For $\eta \approx 0.78$, the diagram shows that the $\mu_6$-C atom has three identical attractors (lone pairs) in the [$\mu_6$-C $\mu_2$-$S_3$]-plane for both complexes $[Re_{12}CS_{17}(CN)_6]^{n-}$ (n = 6, 8). The attractors are directed from the center $\mu_6$-C of the triangle towards its apexes made by $\mu_2$-S atoms to form the angles 120°. Along the triad axis in the plane perpendicular to the [$\mu_6$-C-$\mu_2$-$S_3$] plane there is another region of electron localization ($\eta \approx 0.78$) in the vicinity of $\mu_6$-C that can be also interpreted as a lone pair. The discovered pattern of electron localization suggests that the electron configuration of $\mu_6$-C is made by $sp^2$-hybrids. Since $\eta$ is close to 0.8 at each of the four attractors corresponding to the three $sp^2$ hybrid orbitals and the remaining nonhybrid p orbital, the oxidation state is close to –4 in both complexes.

When the oxidation state of $[Re_{12}CS_{17}(CN)_6]^{n-}$ changes, the only results is the change in the character of the interactions within the systems [$\mu_6$-C$\mu_2$-$S_3$] and [$\mu_6$-C{$Re_6$}$_2$]. For $[Re_{12}CS_{17}(CN)_6]^{6-}$ an unexpected binding region ($\eta \approx 0.3$) is seen between $\mu_6$-C and $\mu_2$-S in the group [$\mu_6$-C$\mu_2$-$S_3$] which is not observed for $[Re_{12}CS_{17}(CN)_6]^{8-}$ (Fig.2 and 3).

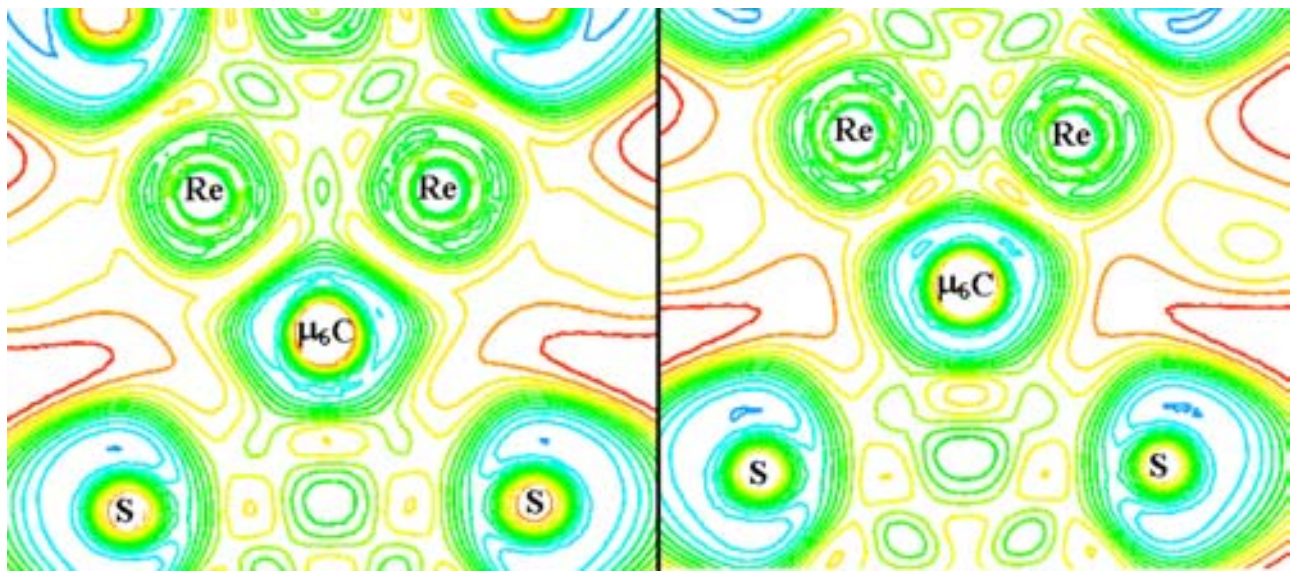

Figure 4. Distribution maps of the same-spin pair density (ELF) for $[Re_{12}CS_{17}(CN)_6]^{6-}$ (left) and $[Re_{12}CS_{17}(CN)_6]^{8-}$ (right). The planes cross the atoms Re-$\mu_6$-C-Re (the rheniums belong to the same fragment [$Re_6$]).

The binding region between $\mu_6$-C and the rheniums in the fragments [$Re_6$] is associated with different levels of electron density localization: ELF ≈ 0.36 for $[Re_{12}CS_{17}(CN)_6]^{6-}$ and ELF ≈ 0.32 for $[Re_{12}CS_{17}(CN)_6]^{8-}$ (Fig.4).

The rhenium atoms inside [$Re_6$] fragments are connected with each other by the regions of electron localization $\eta \approx 0.5$ (green color) which is indicative of shared interactions (covalent bonding [12, 16]) for transition metals, both disynaptic ($\eta \approx 0.42$) and trisynaptic ($\eta \approx 0.36$) attractors being observed. Note that these ELF values do not change together with the oxidation state of $[Re_{12}CS_{17}(CN)_6]^{n-}$.

Hence, the electronic state of $\mu_6$-C in the trigonal-prismatic environment can be interpreted as the one formed by filled $sp^2$ hybrid orbitals and having the oxidation state –4. The interactions between $\mu_6$-C and its environment in $[Re_{12}CS_{17}(CN)_6]^{6-}$ and $[Re_{12}CS_{17}(CN)_6]^{8-}$ can be characterized as ionic and correlation. The unexpected result is a disclosure of long-distance ( ~ 3.454 Å) interactions between $\mu_6$-C and $\mu_2$-S in the [$\mu_6$-$CS_3$] group for $[Re_{12}CS_{17}(CN)_6]^{6-}$ and the absence of such interaction in $[Re_{12}CS_{17}(CN)_6]^{8-}$. Note that a flat trithiocarbonate ligand $[CS_3]^{2-}$ exists in nature but with C-S distances 1.6-1.7 Å [17]. The obtained result of the topological analysis correlates with the sharp change of the distances between the atoms in the vicinity of $\mu_6$-C when the oxidation state of the complexes changes[3]. We hypothesize that the revealed long-distance interaction (the region of electronic density) is responsible for the electrostatic attraction between the rheniums and thereby for the shorter distance (D) in $[Re_{12}CS_{17}(CN)_6]^{6-}$.

We believe that the obtained results are important to the better understanding of the carbon's coordination behavior.

ACKNOWLEDGMENT. The study was supported by State Contract 2007-3-1.3 № 07-01-093 and Russian Foundation for Basic Research (Grants 07-03-00912).

## REFERENCES

1. Zheng, Y.-Q.; von Schnering, H. G.; Chang, J.-H.; *Z. Anorg. Allg. Chem.* 2003, *629*, 1256-1264.
2. Welch, E. J.; Crawford, N. R. M.; Bergman, R. G.; Long, J. R., *J. Am. Chem. Soc.* 2003, *125*, 11464-11465.
3. Mironov, Yu.V.; Naumov, N.G.; Kozlova, S.G.; Kim, Sung-Jin; Fedorov, V.E. *Angew. Chem., Int. Ed.* 2005, *44*, 6867-6871.
4. Amsterdam Density Functional (ADF) program, Release 2006.02; Vrije Universteit: Amsterdam, The Netherlands, 2006.
5. Vosko, S.H.; Wilk, L.; Nusair. M. *Can. J. Phys.* 1980, *58*, 1200-1211.
6. Becke, A.D. *Phys. Rev. A*. 1988, *38*, 3098-3100.
7. Perdew, J.P. *Phys. Rev.B.* 1986, 33, N12, 8822-8824.
8. Van Lenthe, E.; Ehlers, A.E.; Baerends, E.J. *J. Chem. Phys.* 1999, *110*, 8943-8953.
9. Becke, A.D.; Edgecombe, K.E. *J.Chem.Phys*. 1990, *92*, 5387-5403.
10. Savin, A.; Jepsen, O.; Flad, J.; Andersen, O. K.; Preuss, H.; von Schnering, H. G. *Angew. Chem.* 1992, *31*, 187-188.
11. Silvi, B.; Savin, A. Nature, 1994, *371*, 683-686;
12. Berski, S.; Gutsev, G.L.; Mochena, M.D.; Andre´s, J. *J. Phys. Chem. A* 2004, *108,* 6025-6031.
13. Noury, S.; Silvi, B.; Gillespie, R. J. *Inorg. Chem*., 2002, *41* (8) 2164-2172.
14. Kim Ae Ja, Butler L.G. *Inorg.Chem*. 1993, *32*, 178-181.
15. Macchi, P.; Davide M. Proserpio, D.M.; Sironi A. *J. Am. Chem. Soc.* 1998**,** *120,* 13429-13435.
16. Silvi, B.; Gatti, C. *J.Phys.Chem*. A 2000, *104*, 947-953.
17. Simonnet-Je´gat, C.; Cadusseau, E.; Dessapt, R.; Se´cheresse, F. *Inorg. Chem.* 1999, *38,* 2335-2339.

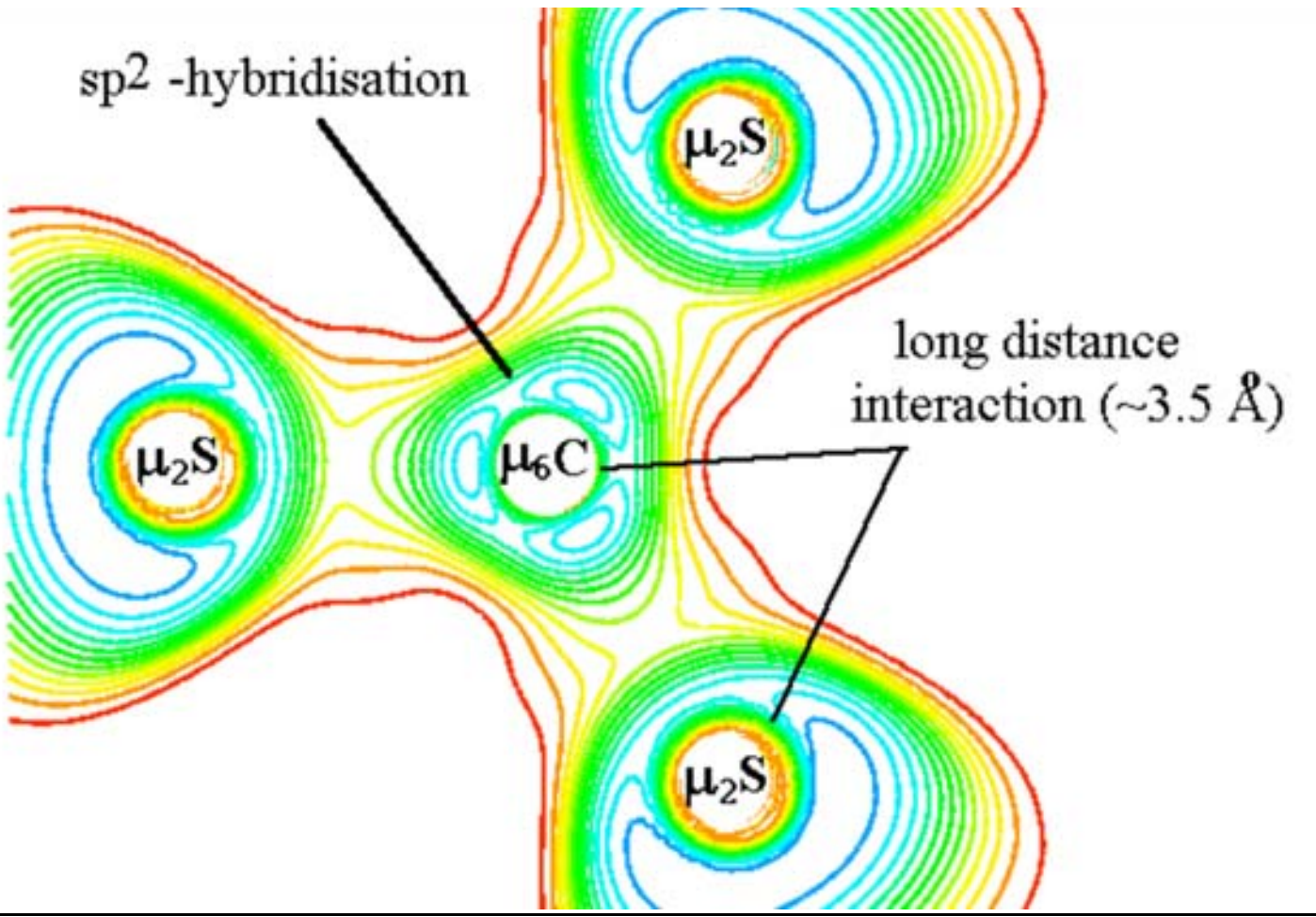

**ABSTRACT FOR WEB PUBLICATION**

The electronic state of carbon $\mu_6$-C in trigonal prismatic environment of rheniums in $[Re_{12}\mu_6\text{-}CS_{17}(CN)_6]^{n-}$ complexes (n = 6, 8) was studied with electron localization function (ELF) and was shown to be characterized by $sp^2$-hybridisation and oxidation state –4. The character of multicentered interactions in $[\mu_6\text{-C–}\mu_2\text{-}S_3]$ and $[\mu_6\text{-C–}\{Re_6\}_2]$ groups changes together with the oxidation state. Unexpected long-distance interaction was obtained between $\mu_6$-C and $\mu_2$-S in the group $[\mu_6\text{-C–}\mu_2\text{-}S_3]$ for $[Re_{12}\mu_6\text{-}CS_{17}(CN)_6]^{6-}$ complex.